\documentstyle[twocolumn,prl,aps,epsfig,amssymb]{revtex}
\tighten
\let\jnfont=\rm
\def\NPB#1,{{\jnfont Nucl.\ Phys.\ B }{\bf #1},}
\def\PLB#1,{{\jnfont Phys.\ Lett.\ B }{\bf #1},}
\def\EPJC#1,{{\jnfont Eur.\ Phys.\ Jour.\ C }{\bf #1},}
\def\PRD#1,{{\jnfont Phys.\ Rev.\ D }{\bf #1},}
\def\PRL#1,{{\jnfont Phys.\ Rev.\ Lett.\ }{\bf #1},}
\def\MPLA#1,{{\jnfont Mod.\ Phys.\ Lett.\ A }{\bf #1},}
\def\JPG#1,{{\jnfont J.\ Phys.\ G}{\bf #1},}
\def\CTP#1,{{\jnfont Commun.\ Theor.\ Phys.\ }{\bf #1},}

\begin{document}

\preprint{\parbox{2.0in}{\noindent hep-ph/0608315}}

\title{Constraints on Hybrid Inflation from Flat Directions in Supersymmetry}

\author{\ \\[2mm]  Fuqiang Xu $^1$,  Jin Min Yang $^{2,1}$}

\address{ \ \\[2mm]
{\it $^1$ Institute of Theoretical Physics, Academia Sinica, Beijing 100080, China} \\ [2mm]
{\it $^2$ CCAST (World Laboratory), P.O.Box 8730, Beijing 100080, China}}

\maketitle

\begin{abstract}
We examine the constraints on F-term hybrid inflation by considering the
flat directions in the Minimal Supersymmetric Standard Model (MSSM).
We find that some coupling terms between the flat direction fields
and the field which dominates the energy density during inflation
are quite dangerous and can cause the no-exit of hybrid inflation
even if their coupling strength is suppressed by Planck scale.
Such couplings must be forbidden by imposing some symmetry for
a successful F-term hybrid inflation.
At the same time, we find that in the D-term inflation
these couplings can be avoided naturally. Further, given the tachyonic preheating,
we discuss the feasibility of Affleck-Dine baryogenesis after
the F-term and D-term inflations.
\end{abstract}


{\em Introduction:~}
Hybrid inflation is now the standard paradigm of inflation and has
proved very fruitful since it was introduced \cite{hybrid}.
In the models of hybrid inflation, most of the energy density is
provided by an auxiliary field $\sigma$ instead of the slowly rolling
inflaton field $\Phi$.
When $\sigma$ falls below a critical value $\sigma_c$, the inflation
ends, which gives a natural exit of inflation.
But when a particle physics model is tried for hybrid inflation,
e.g., F-term hybrid inflation, there always exist a mass term of inflaton
which can be of the same order as the Hubble scale and destroy the slow-rolling
condition (the so-called $\eta$ problem).
So far various approaches have been proposed to avoid this $\eta$ problem \cite{lythriotto}.

The flat directions, abundant in the MSSM, have many distinctive
features and may thus play an important role in the early universe.
Such flat directions are protected from perturbative quantum
corrections and only lifted by the soft breaking terms which are
insignificant in the early universe. During the inflation these flat
directions can get large vacuum expectation values (vev) which
naturally make an initial condition for many phenomena. From the
potential of a flat direction $\varphi$ during inflation \cite{drt}
\begin{eqnarray}
V (\varphi)&=& (m_{0}^{2}+c_{H}H_I^{2}) |\varphi|^{2}
+\frac{A\lambda H_I\varphi^n e^{in\theta_\varphi}+h.c.}{n M^{n-3}} \nonumber\\
& &+ \lambda^{2}\frac{|\varphi|^{2(n-1)}}{M^{2(n-3)}} ,
\label{eqb}
\end{eqnarray}
we get the large vevs
\begin{equation}
 \varphi_{0} \sim(H_{I}M^{n-3}/\lambda)^{1/n-2} . \label{eqa}
 \end{equation}
Here $n$ is an integer ($\geq 4$),  $H_I$ is the Hubble parameter
during inflation, $m_0$ is the soft mass term, $c_H$ ($<0$) and $A$
are constants of ${\cal O}(1)$, and $M$ is usually the Planck scale
\footnote{In fact if $M$ is the GUT or smaller scale, the problem of
no preheating in chaotic inflation (mentioned below) will be
alleviated.  The same thing will happen in hybrid inflation
considered in our work below. But to be conservative, we consider
the worst case where $M$ is taken as the Planck scale.}. Note that
the large vevs of the flat directions can also cause some problems
in various inflation models \cite{Pilaftsis,Volansky}, especially
they can kinematically block the resonance preheating in chaotic
inflation due to the invalidation of the non-adiabaticity condition
\cite{alla}.

As noticed in \cite{alla}, the large vevs of the  flat directions
can also cause no exit of hybrid inflation. In this work we will
scrutinize this problem and find that such a problem can be caused
by some coupling terms between the flat direction fields and the
field which dominates the energy density during inflation. Such
couplings must be forbidden by imposing some symmetry for a
successful F-term hybrid inflation. Then we point out that in the
D-term inflation these couplings can be avoided naturally and thus
the problem of no-exit of hybrid inflation does not exist, and
further the no-preheating problem can also be avoided.

We will also discuss the feasibility of Affleck-Dine baryogenesis
for both F-term and D-term inflations, given the tachyonic
preheating. It is well known that after chaotic inflation ends, the
process of resonance preheating occur at once \cite{resonance}. This
preheating leads to the large oscillation of the fields which
couples with inflaton, and can make the symmetries restored for a
moment which can make the baryon number produced after chaotic
inflation \cite{alri}. Noticing that similar processes may occur
during tachyonic preheating \cite{tachyonic} after hybrid inflation,
we find that it is possible the baryon number produced in these
phase is copious due to the large amplitude of inflaton.
\vspace*{0.5cm}

{\em Flat directions constraints in F-term hybrid inflation:~}
So far in the literature the couplings of inflaton to
matter fields have not been intensively studied and only some toy models
have been considered which have no relevance to SM particles \cite{resonance,8989}.
In \cite{mazu} the importance of gauge invariance was first highlighted
and recently the strength of the couplings of inflaton to matter fields
was studied \cite{alla}.
In the following we will examine the couplings between the
MSSM fields and the auxiliary field dominating the energy
density where the gauge symmetry is also important.

In the simplest supersymmetric hybrid inflation model,
the superpotential contains the terms
\begin{equation}
W \supset g \hat\phi(\hat\sigma^{2}-\sigma_{0}^{2}) ,
\label{eqc}
\end{equation}
where $g$ is a coupling constant,  $\hat\phi$ is the inflaton
superfield and $\hat\sigma$ is the superfield containing the scalar
field $\sigma$. Large vevs of flat directions can induce a mass
$\lambda_1\varphi_{0}$ for the $\sigma$ field if we assume the
existence of the $\lambda_1^2|\varphi|^2|\sigma|^2$ interaction with
$\lambda_1$ being a coupling constant (such a term is assumed to have
the same sign as the interaction terms from the superpotential $W$). Then if
\begin{equation} \label{condition}
\lambda_1\varphi_{0} > \sqrt{2}g\sigma_{0} ,
\end{equation}
the mass-square of $\sigma$ will remain positive even for $\langle \phi \rangle < \phi_{c}$
($\phi_{c}$ is the critical value at which the hybrid inflation naturally exits),
which leads to $\langle \sigma \rangle =0$ and no tachyonic preheating and no
exit from hybrid inflation.

In fact, the above problem always exists in hybrid inflation if
$\lambda_1 > \sqrt{2}g$, which can be seen clearly by replacing
$H_{I}=\sqrt{V_0/3M_P^2}$ with $V_0=g^2 \sigma_{0}^4$ into
Eq.(\ref{eqa}). We find that if we ignore the coupling constants in
Eq.(\ref{eqb}), then we have $\varphi_0\ge \sigma_0$, where
$\sigma_0$ decides the dominant energy density in hybrid inflation
even for $n=4$.

Then a natural idea for solving this problem is to suppress the
dangerous large coupling $\lambda_1$ to invalidate
Eq.(\ref{condition}). But it is difficult to find out a way in
generic hybrid inflation. For example, let us consider the matter
fields contained in the MSSM flat directions besides the two scalar
fields $\sigma$ and $\phi$ in hybrid inflation. With the addition of
$\sigma$ field, the superpotential is obtained by multiplying the
$\sigma$ superfield with the following MSSM gauge invariant terms
 \begin{equation}
 H_uH_d \ , \ H_uL ,
 \end{equation}
and
 \begin{equation}
 H_u Qu \ , \ H_dLe \ , H_dLe \ , \ QLd \ , \ udd \ , \ LLe ,
 \label{eqf}
 \end{equation}
where $H_u$ and $H_d$ are the two Higgs doublets, $L$ and $Q$
denote respectively a doublet of lepton and quark,
and $e$, $u$ and $d$ denote respectively a singlet of lepton,
up-quark and down-quark.
Here we drop out the higher order gauge-invariant terms, which are
negligible, as shown in our following analysis. We will label a
generic MSSM flat direction by $\varphi$. Examples of such flat
directions are given by $LH_u \sim \varphi^2$ in terms of
doublet components $L=(\varphi, 0)$ and $H_u=(0, \varphi)$, and by
$Q_1L_1 d_2 \sim \varphi^3$. Then we obtain interactions
like
\begin{equation} \label{eq4}
\lambda^2_1 |\varphi|^2 \sigma^2 + \lambda^2_2 |\varphi|^4 \sigma^2/M_P^2
\end{equation}
where $\lambda_i (i=1,2)$ are coupling constants.

Let us consider the renormalizable terms, i.e. $\sigma H_uH_d$ and
$\sigma H_u L$. It is natural to assume their coupling constants to
be ${\cal O}(1)$. On the other hand, we know that these gauge
invariant terms correspond to the D-flat directions of the MSSM
\cite{gkm}. These terms $\sigma H_uH_d$ and $\sigma H_u L$ can lead
to $\lambda_{1,2} \sim {\cal O}(1)$. Since $\varphi_0\ge \sigma_0$
even for $n=4$ (D-flat $H_uH_d$ and $H_u L$ directions are lifted by
$n=4$ non-renormalizable terms), it is impossible to avoid
Eq.(\ref{eqf}) unless $g \gg {\cal O}(1)$.

Actually, the coupling terms in Eq.(\ref{eq4}) with even a small
$\lambda_i$ cannot be neglected because the vevs of the flat
directions which are lifted by a large $n$ ($n$ can be as large as 9
\cite{gkm}) are much larger than $\sigma_0$. From some calculations
we find that there are two other dangerous flat directions: $LLe$
and $udd$, which are lifted up by $n=6$ non-renormalizable terms.

Some global symmetry \footnote{Note that unlike a discrete symmetry
\cite{discr}, such a global symmetry may have a potential problem
since it may be violated by quantum gravity effects \cite{grav}.}
like $R$-symmetry $U(1)_R$ \cite{R-symmetry} must be imposed to
forbid the 4 dangerous terms $H_uH_d$, $H_uL$, $LLe$ and $udd$ in
Eq.(\ref{eqf}). If we use $R$-symmetry, we must make special
assignment for the $R$-charges of involved superfields, e.g., the
$R$-charge must be zero for $\sigma$ field as can be seen from
Eq.(\ref{eqc}).

Now we discuss the feasibility of Affleck-Dine baryogenesis during
tachyonic preheating after the F-term hybrid inflation. After the
inflation the $A$-term in the potential in Eq.(\ref{eqb}) becomes
small as the Hubble parameter $H$ decreases, and there will appear a
CP-violating interaction \cite{alri}
\begin{equation}
c(\phi^2/M_P^{m-2})\varphi^m + h.c.
\end{equation}
During tachyonic preheating, the large oscillation
of inflaton $\phi$ can make the flat directions
go to zero and
\begin{equation}
\phi^2 \sim \langle \phi^2 \rangle e^{i2\theta_{\phi}} .
\end{equation}
Through Affleck-Dine mechanism, this process will
produce baryon number
\begin{equation}
n_B \sim 2|\varphi_0|^2 \dot\theta_\varphi .
\end{equation}
The initial value of $\theta_\varphi$ is determined by the $H_I$-dependent
$A$-term in Eq.(\ref{eqb}).
Such baryon number will be released in the ensuing decay of $\varphi$.
Because $\varphi_0$ is so large, this baryon number generated during the
tachyonic preheating after hybird inflation cannot be neglected.
This is quite similar to the case of resonance
preheating after chaotic inflation considered in \cite{alri}.
\vspace*{0.5cm}

{\em D-term inflation and its consequence:~} In the preceding section
we have shown that it is not easy to obtain a negative mass-square
term due to the large vevs of flat directions, which will lead to
no elegant exit of hybrid inflation in the F-term inflation.
Here we show that such a problem can be solved naturally in D-term inflation.

D-term inflation can preserve the flat directions of global
supersymmetry and, in particular, keep the inflation potential flat
provided that one of the contributions to the potential $V_D$
contains a Fayet-Iliopoulos term as in Eq. (\ref{eqd}). This was
first pointed out in \cite{stewart} and significantly improved in
\cite{bdh}.

Consider a toy model of D-term inflation, whose field content
includes a inflaton chiral superfield $\hat\phi$ and auxiliary superfields
$\hat \sigma_{\pm}$ with charges $\pm 1$ under an anomalous $U(1)_X$
symmetry. From the superpotential
 \begin{eqnarray}
W = c\hat\phi \hat\sigma_+ \hat\sigma_- ,
 \end{eqnarray}
and the minimal Kahler potential,
we obtain the tree-level scalar potential
 \begin{eqnarray}
 V &=& |c|^2 (|\sigma_+ \sigma_-|^2 + |\phi \sigma_+|^2 + |\phi \sigma_-|^2)\nonumber \\
   & & + \frac{g^2}{2}(|\sigma_+|^2-|\sigma_-|^2+\xi^2)^2
 \label{eqd}
 \end{eqnarray}
Here $c$ is a coupling constant and the Fayet-Iliopoulos term $\xi^2$
is assumed to be positive.
The role of $\sigma$ field in Eq. (\ref{eqc}) is now replaced by the $\sigma_-$
field in this model.

Now we argue that in the D-term inflation the no-exit problem happened
in generic F-term inflation discussed in the preceding section
does not exist.  The magnitude of the vevs of
the MSSM flat directions are primarily fixed by higher
dimension operators in the Kahler potential that couple the flat
direction to other fields like
\begin{eqnarray}
 \bigtriangleup \mathcal{L} &=& \int d^4 \theta (c_1|\hat\phi|^2 + c_2|\hat\sigma_+|^2 +
 c_3|\hat\sigma_-|^2) \frac{\varphi^+\varphi}{M^2} ,
 \label{eqe}
 \end{eqnarray}
where $c_i (i=1,2,3)$ are coupling constants. These induced mass
terms for flat direction $\varphi$ always vanish because $\langle \sigma_- \rangle
=0$ and $\langle \sigma_+ \rangle =0$ during D-term inflation. Here
the zero vev of $\sigma_{\pm}$ can be obtained by minimizing the potential
of Eq. (\ref{eqd}) when $\phi > \phi_c = g\xi/c$ with
$\phi_c$ being the critical value at which inflation ends. Then
the flat directions can only get vev after inflation where $U(1)_X$
is broken and we get
 \begin{eqnarray}
 \langle \sigma_+ \rangle  =0,\
 \langle \sigma_- \rangle \sim \xi,\
 \langle F_{\sigma_+} \rangle \sim \xi\phi .
 \end{eqnarray}
After the quick tachyonic preheating, the inflaton $\phi$ fall down
to its global minimum and equal to zero. So the $\langle
F_{\sigma_+} \rangle \sim 0$, but there always exist the kinetic
energies in Eq. (\ref{eqe}) which will lead to the negative mass-square
term of flat directions fields.
Of course, such delayed appearance of the vevs of flat directions
will not cause the problem of no-exit of inflation.

On the other hand, as pointed in \cite{alla}, the large vev of flat
directions will prohibit the tachyonic preheating in F-term
inflation. However, we should point out that this problem does not
happen in D-term inflation because the direct coupling terms between
flat directions and $\sigma_{\pm}$ are forbidden by $U(1)_X$ gauge
symmetry. Then the preheating will proceed despite of the large vevs
of the flat directions. The ensuing Affleck-Dine baryogenesis can
occur as indicated in \cite{kmar} after tachyonic preheating, but
the decays of inflaton and $\sigma_{\pm}$ must be quite different
with those in no-preheating case. \vspace*{0.5cm}

{\em Conclusion:~}
We discussed the effects of large vevs of MSSM flat directions
in hybrid inflation in detail. We have shown that some dangerous coupling
terms between the flat direction fields and the field which dominate the
energy density during inflation must be forbidden for a successful inflation
even if their coupling constants are small.
Such couplings must be forbidden by imposing some symmetry for
a successful F-term hybrid inflation.
At the same time, we found  that in the D-term inflation these couplings
can be avoided naturally.
Further, given the tachyonic preheating, we discussed the feasibility of
Affleck-Dine baryogenesis after the F-term and D-term inflations.
\vspace*{0.5cm}

This work was supported in part by National Natural Science
Foundation of China (NNSFC) under number No. 10475107.

\end{document}